\documentclass{article}
\usepackage[utf8]{inputenc}
\usepackage[top=3cm, bottom=3cm, left=3cm,right=3cm]{geometry}
\usepackage{natbib}
\setcitestyle{aysep={}} 
\usepackage{graphicx}
\usepackage{color}
\usepackage{amsfonts, amsmath}
\usepackage{setspace}
\usepackage{amsmath}
\usepackage{mathtools}
\usepackage{bbm}
\usepackage{float}
\usepackage{hyperref}
\usepackage[super]{nth}

\usepackage{lineno}
\let\oldalign\align
\let\oldendalign\endalign
\renewenvironment{align}{\linenomathNonumbers\oldalign}{\oldendalign\endlinenomath}

\makeatletter
\newcommand{\customlabel}[2]{
\protected@write \@auxout {}{\string \newlabel {#1}{{#2}{}}}}
\makeatother

\renewcommand{\eqref}[1]{\ref{#1}}

\begin{document}
{\Large Ancestral process for infectious disease outbreaks with superspreading}


\vspace*{2cm}
Xavier Didelot$^{1,2,*}$, David Helekal$^{3}$, Ian Roberts$^{2}$

\vspace*{2cm}
$^1$ School of Life Sciences, University of Warwick, Coventry, United Kingdom\\
$^2$ Department of Statistics, University of Warwick, Coventry, United Kingdom\\
$^3$ Department of Immunology and Infectious Diseases, Harvard T. H. Chan School of Public Health, Boston, Massachusetts, USA\\
$^*$ Corresponding author. Tel: 0044 (0)2476 572827. Email: \verb+xavier.didelot@warwick.ac.uk+

\vspace*{2cm}
Running title: Ancestry for outbreaks with superspreading

\vspace*{2cm}

\newpage
\section*{Abstract}

When an infectious disease outbreak is of a relatively small size, describing the ancestry
of a sample of infected individuals is difficult because most ancestral models assume
large population sizes. Given a set of infected individuals, 
we show that it is possible to express exactly
the probability that they have the same infector, either inclusively
(so that other individuals may have the same infector too) or exclusively 
(so that they may not).
To compute these probabilities requires knowledge of the offspring distribution, 
which determines how many infections each infected individual causes. 
We consider transmission both without and with superspreading, in the form of a 
Poisson and a Negative-Binomial offspring distribution, respectively. 
We show how our results can be incorporated into a new lambda-coalescent model
which allows multiple lineages to coalesce together. 
We call this new model the omega-coalescent, we compare it
with previously proposed alternatives, and advocate its use
in future studies of infectious disease outbreaks.

\newpage
\section{Introduction}

An outbreak of an infectious disease typically starts when a single or a small number
of infected individuals appear within a susceptible population. Each infected individual
may come in contact and transmit the disease to each of the susceptible individuals, who will then
become infected in their turn and spread the disease further. Most mathematical models of infectious diseases
describe situations where the disease is at an equilibrium, when the number
of infected individuals is high and/or with a significant part of the population already infected
\citep{Anderson1991,keeling2008modeling}. 
Here however we focus on the early stages of an epidemic, where the number of 
infected individuals is small and the number of susceptibles comparatively high and constant.
In this situation it is useful to consider the number of new infections that each infected
individual is likely to cause, and the probabilistic distribution for this number is often called
the offspring distribution \citep{Grassly2008}. 
The mean of the offspring distribution is called the basic
reproduction number $R_0$ and has been given much attention especially since
it determines how likely the outbreak is to spread, and how much effort would be needed
to bring it under control \citep{fraserFactorsThatMake2004,fergusonStrategiesMitigatingInfluenza2006}. 

If we consider that all
individuals are infectious for the same duration and with the same transmission rate,
the offspring distribution is Poisson distributed with mean $R_0$, in which case the variance
of the offspring distribution is also $R_0$. We would then say that there is no transmission
heterogeneity. However, in practice there are many reasons why this may not be the case,
with some individuals being infectious for longer than others, or being more infectious than others, or 
having more frequent contacts with susceptibles, or being less symptomatic and therefore less likely
to reduce contact numbers, etc. All these factors cause the offspring distribution to be more 
dispersed than it would otherwise be, that is to have a variance greater than its mean $R_0$. 
A frequent choice to capture this overdispersion is to model the offspring distribution
using a Negative-Binomial distribution with mean $R_0$ and dispersion parameter $r$
\citep{Lloyd-Smith2005,Grassly2008}. When $r$ is close to zero the variance is high compared to the mean,
whereas when $r$ is high the variance becomes close to the mean. This transmission heterogeneity
is often called superspreading, although this is perhaps misleading
as it is the rule rather than the exception of how infectious diseases spread. Superspreading has
indeed been described in many diseases \citep{woolhouseHeterogeneitiesTransmissionInfectious1997,
steinSuperspreadersInfectiousDiseases2011,kucharskiRoleSuperspreadingMiddle2015,
wangSuperspreadingHeterogeneityTransmission2021}, 
and most recently for 
SARS-CoV-2 \citep{Wang2020,lemieuxPhylogeneticAnalysisSARSCoV22021,gomez-carballaSuperspreadingEmergenceCOVID192021,duSystematicReviewMetaanalyses2022}.

As an outbreak unfolds forward-in-time, a transmission tree is generated representing who-infected-whom,
in which each node is an infected individual and points towards a number of nodes distributed according
to the offspring distribution. Here we consider the reverse problem of the transmission ancestry, going backward-in-time, from a sample of infected individuals, until reaching the last common 
transmission ancestor of the whole sample. Given a set of $n$ sampled individuals,
we show how to calculate the probability that a given subset of size $k$ have the same infector,
either inclusively (so that the remaining $n-k$ may also have the same infector or not)
or exclusively (so that none of the remaining $n-k$ have the same infector). We start
by considering the general case of an offspring distribution with arbitrary form, 
and then the specific cases of 
offspring distributions that follow a Poisson and a Negative-Binomial distribution.
The main novelty of our approach is that we consider that the overall population size is small,
but we show that in the limit where the population size is large, our results agree with several previous
studies \citep{Volz2012a,koelleRatesCoalescenceCommon2012,Fraser2017}.
Finally, we show how our results can be incorporated into a new lambda-coalescent model 
\citep{pitmanCoalescentsMultipleCollisions1999,sagitovGeneralCoalescentAsynchronous1999,
donnellyParticleRepresentationsMeasureValued1999} and compare it with previously
proposed models.

\section{General offspring distribution case}

Let time be measured in discrete units and denoted $t$. Each discrete value of $t$ corresponds to a unique non-overlapping 
generation of infected individuals, so that individuals infected at $t$ have offspring at $t+1$, etc. 
Let $N_t$ denote the number of infectious individuals at time $t$. Each of them creates a number $s_{t,i}$ of secondary infections at time $t+1$, following the offspring distribution $\alpha_t(s)$. The mean of this distribution is the basic reproduction number $R_t$ and the variance is $V_t$. The total number of infected individuals at time $t+1$ is given by:

\begin{equation}
N_{t+1}=\sum_{i=1}^{N_t} s_{t,i}
\label{eq:summation}
\end{equation}

\subsection{Inclusive coalescence probability}

We define the inclusive coalescence probability $p_{k,t}(N_t, N_{t+1})$ as the probability that a specific set of $k$ individuals from generation $t+1$ have the same infector in generation $t$, conditional on population sizes $N_t$ and $N_{t+1}$.
Given full information about offspring counts from individuals in generation $t$, ${\bf s}_t = (s_{t,1}, \dots s_{t, N_t})$, we have:

{\allowdisplaybreaks
	\begin{align}
		p_{k,t}({\bf s}_t, N_t)
			& = \sum_{i=1}^{N_t} \frac{\binom{s_{t,i}}{k}}{\binom{N_{t+1}}{k}} \nonumber\\
			 &= \sum_{i=1}^{N_t} \frac{s_{t,i}!}{(s_{t,i}-k)!} \frac{(N_{t+1}-k)!}{N_{t+1}!} 
	\end{align}
}

Full information $\{s_{t,i}\}$ yields the population size $N_{t+1}$ as shown in Equation \eqref{eq:summation}, 
but this is not available in practice.
We can instead express the inclusive coalescence probability conditioning on the next population size $N_{t+1}$ by summing over possible offspring counts ${\bf s}_t = (s_{t,1}, \dots s_{t, N_t})$ conditional on the total generation size.
Let $S_t^{-(1)} = (S_{t,2}, \dots, S_{t, N_t})$:

{\allowdisplaybreaks
	\begin{align}
	p_{k,t}(N_t, N_{t+1})
		& = \sum_{{\bf s}_t \in \mathbb{N}_0^{N_t}} \mathbb{P} \bigg[{\bf S}_t = {\bf s}_t \bigg| \sum_{i=1}^{N_t} S_{t,i} = N_{t+1} \bigg] p_{k,t}({\bf s}_t, N_t) \nonumber\\
		& = \sum_{{\bf s}_t \in \mathbb{N}_0^{N_t}} \mathbb{P} \bigg[{\bf S}_t = {\bf s}_t \bigg| \sum_{i=1}^{N_t} S_{t,i} = N_{t+1} \bigg] \sum_{i=1}^{N_t} \frac{\binom{s_{t,i}}{k}}{\binom{N_{t+1}}{k}} \nonumber\\
		& = \sum_{i=1}^{N_t} \sum_{{\bf s}_t \in \mathbb{N}_0^{N_t}}\frac{\binom{s_{t,i}}{k}}{\binom{N_{t+1}}{k}} \mathbb{P} \bigg[S_{t,1} = s_{t,1}, {\bf S}_t^{-(1)} = {\bf s}_t^{-(1)} \bigg| \sum_{i=1}^{N_t} S_{t,i} = N_{t+1} \bigg] \nonumber\\
		& = \frac{N_t}{\binom{N_{t+1}}{k}} 
		\sum_{{\bf s}_t \in \mathbb{N}_0^{N_t}} \binom{s_{t,1}}{k} \mathbb{P} \bigg[S_{t,1} = s_{t,1} \bigg| \sum_{i=1}^{N_t} S_{t,i} = N_{t+1} \bigg] \nonumber \\
			& \phantom{=}\qquad \times \mathbb{P} \bigg[{\bf S}_t^{-(1)} = {\bf s}_t^{-(1)} \bigg| S_{t,1} = s_{t,1}, \sum_{i=1}^{N_t} S_{t,i} = N_{t+1} \bigg] \nonumber\\
		& = \frac{N_t}{\binom{N_{t+1}}{k}} \sum_{s_{t,1} = 0}^{N_{t+1}} \binom{s_{t,1}}{k} \mathbb{P} \bigg[S_{t,1} = s_{t,1} \bigg| \sum_{i=1}^{N_t} S_{t,i} = N_{t+1} \bigg] \nonumber \\
			& \phantom{=}\qquad \times \underbrace{
				\sum_{{\bf s}_t^{-(1)} \in \mathbb{N}_0^{N_t - 1}} \mathbb{P} \bigg[{\bf S}_t^{-(1)} = {\bf s}_t^{-(1)} \bigg| \sum_{i=2}^{N_t} S_{t,i} = N_{t+1} - s_{1,t} \bigg]}_{=1} \nonumber\\
		& = \frac{N_t}{\binom{N_{t+1}}{k}} \mathbb{E}\bigg[ \binom{S_{t,1}}{k} \bigg| \sum_{i=1}^{N_t} S_{t,i} = N_{t+1} \bigg] \nonumber\\
		& = N_t \frac{(N_{t+1} - k)!}{N_{t+1}!} \mathbb{E} \bigg[ \frac{S_{t,1}!}{(S_{t,1} - k)!} \bigg | \sum_{i=1}^{N_t} S_{t,i} = N_{t+1} \bigg] \label{eq:GeneralInclusiveProb}
	\end{align}
}

The $k$-th falling factorial moments $\mathbb{E} \big[ \frac{S_{t,1}!}{(S_{t,1} - k)!} \big | \sum_{i=1}^{N_t} S_{t,i} = N_{t+1} \big]$ in Equation \eqref{eq:GeneralInclusiveProb} can be readily obtained by differentiating the probability generating function of $S_{t,1} | (\sum_{i=1}^{N_t} S_{t,i} = N_{t+1})$.

\subsection{Exclusive coalescence probability}

Generally, we observe a sample of individuals from each generation rather than the entire population.
In this case, we are interested in the exclusive coalescence probability $p_{n,k,t}(N_t, N_{t+1})$ that a specific subset of $k$ individuals amongst $n$ sampled individuals arose from a common infector one generation in the past given knowledge of the total population sizes $N_t$ and $N_{t+1}$.
Let us first assume full knowledge about offspring counts of the individuals at time $N_t$ amongst the sample at time $N_{t+1}$, namely ${\bf x}_t = (x_{t,1}, \dots, x_{t,N_t})$ such that $x_{t,1}+...+x_{t,N_t}=n$. 
Note that $X_{t,i}$ does not follow the same offspring distribution as $S_{t,i}$. We have:
	\begin{align}
		p_{n,k,t}({\bf x}_t, N_t)
			& = \sum_{i=1}^{N_t} \frac{\binom{x_{t,i}}{k}}{\binom{n}{k}} \mathbb{I} \{ x_{t,i} = k \}\nonumber\\
			& = \sum_{i=1}^{N_t} \frac{x_{t,i}!}{(x_{t,i} - k)!} \frac{(n-k)!}{n!} \mathbb{I} \{ x_{t,i} = k \}
	\end{align}

Similarly to the inclusive coalescence probability in Equation \eqref{eq:GeneralInclusiveProb}, we can use this to evaluate the exclusive probability given $N_t$ and $N_{t+1}$ by summing over possible parent offspring configurations (for $k \leq n$):

{\allowdisplaybreaks
	\begin{align}
		p_{n,k,t}(N_t, N_{t+1})
			& = \sum_{{\bf x}_t \in \mathbb{N}_0^{N_t}} \mathbb{P} \bigg[ {\bf X}_t = {\bf x}_t \bigg | \sum_{i=1}^{n} X_{t,i} = n \bigg] p_{n,k,t} ({\bf x}_t, N_t) \nonumber\\
			& = \sum_{{\bf x}_t \in \mathbb{N}_0^{N_t}} \mathbb{P} \bigg[ {\bf X}_t = {\bf x}_t \bigg | \sum_{i=1}^{n} X_{t,i} = n \bigg] \sum_{i=1}^{N_t} \frac{\binom{x_{t,i}}{k}}{\binom{n}{k}} \mathbb{I} \{ x_{t,i} = k \} \nonumber\\
			& = \frac{N_t}{\binom{n}{k}} \sum_{{\bf x}_t \in \mathbb{N}_0^{N_t}} \binom{x_{t,1}}{k} \mathbb{P} \bigg[ {\bf X}_t = {\bf x}_t \bigg| \sum_{i=1}^{N_t} X_{t,i} = n \bigg] \mathbb{I} \{x_{t,1} = k \} \nonumber\\
			& = \frac{N_t}{\binom{n}{k}}
				\sum_{{\bf x}_t^{-(1)} \in \mathbb{N}_0^{N_t - 1}} \binom{k}{k} \mathbb{P}\bigg[X_{t,1} = k, {\bf X}_t^{-(1)} = {\bf x}_t^{-(1)} \bigg| \sum_{i=1}^{N_t} X_{t,i} = n \bigg] \nonumber\\
			& = \frac{N_t}{\binom{n}{k}} \mathbb{P}[X_{t,1} = k \bigg| \sum_{i=1}^{N_t} X_{t,i} = n \bigg]
				\underbrace{\sum_{{\bf x}_t^{-(1)} \in \mathbb{N}_0^{N_t - 1}} \mathbb{P}\bigg[{\bf X}_t^{-(1)} = {\bf x}_t^{-(1)} \bigg| \sum_{i=1}^{N_t} X_{t,i} = n, X_{t,1} = k \bigg]}_{=1} \nonumber\\
			& = \frac{N_t}{\binom{n}{k}} \mathbb{P} \bigg[ X_{t,1} = k \bigg| \sum_{i=1}^{N_t} X_{t,i} = n \bigg] \label{eq:GeneralExclusiveProb}
	\end{align}
	}


\subsection{Complementarity of exclusive coalescence probabilities}

If we consider one of the lines observed amongst a set of $n$, it can either remain uncoalesced
with probability $p_{n,1,t}(N_t, N_{t+1})$ or coalesce in an event of size $k$ with probability $p_{n,k,t}(N_t, N_{t+1})$ with any set of $k-1$ lines among the $n-1$ other lines, leading to the following complementarity equation: 

\begin{equation}
\sum_{k=1}^n \binom{n-1}{k-1} p_{n,k,t}(N_t, N_{t+1}) =1
\end{equation}

We can show that it is indeed satisfied by the formula in Equation \eqref{eq:GeneralExclusiveProb}:

	\begin{align}
		\sum_{k=1}^n \binom{n-1}{k-1} p_{n,k,t}(N_t, N_{t+1})
			& = \sum_{k=1}^n \binom{n-1}{k-1} \frac{N_t}{\binom{n}{k}} \mathbb{P}\left[X_1 = k \bigg| \sum_{i=1}^{N_t} X_i = n\right] \nonumber\\
			& = \sum_{k=1}^n N_t \frac{k}{n} \mathbb{P}\left[X_1=k \bigg| \sum_{i=1}^{N_t} X_i = n\right] \nonumber\\
			& = \frac{N_t}{n} \sum_{k=0}^n k \mathbb{P}\left[X_1 = k \bigg| \sum_{i=1}^{N_t} X_i = n\right] 
			\nonumber\\
			& = \frac{N_t}{n} \mathbb{E}\left[X_1 \bigg| \sum_{i=1}^{N_t} X_i = n \right] \nonumber\\
			& = \frac{1}{n} \sum_{i=1}^{N_t} \mathbb{E}\left[X_i \bigg| \sum_{i=1}^{N_t} X_i = n\right] 
			\nonumber\\
			& = \frac{1}{n} \mathbb{E}\left[ \sum_{i=1}^{N_t} X_i \bigg| \sum_{i=1}^{N_t} X_i = n\right] \nonumber\\
			& = 1
	\end{align}
	
\section{Poisson offspring distribution case}

In this section we consider that the offspring distribution is $\alpha_t = \text{Poisson}(R_t)$.
In this case, we have: 
	\begin{equation}
		\sum_{i=1}^{N_t} S_{t,i} \sim \text{Poisson}(N_t R_t)
	\end{equation}
and the conditional distribution:
{\allowdisplaybreaks
	\begin{align}
	\allowdisplaybreaks
		\mathbb{P}\bigg[S_{t,1} = s \bigg| \sum_{i=1}^{N_t} S_{t,i} = N_{t+1} \bigg]
			& = \frac{\displaystyle\mathbb{P}\bigg[S_{t,1} = s, \sum_{i=1}^{N_t} S_{t,i} = N_{t+1} \bigg]}{\displaystyle\mathbb{P}\bigg[\sum_{i=1}^{N_t} S_{t,i} = N_{t+1} \bigg]} \nonumber\\
			& = \frac{\displaystyle\alpha_t (s) \, \mathbb{P}\bigg[ \sum_{i=2}^{N_t} S_{t,i} = N_{t+1} - s \bigg]}{\displaystyle\mathbb{P}\bigg[\sum_{i=1}^{N_t} S_{t,i} = N_{t+1} \bigg]} \nonumber\\
			& = \frac{\displaystyle\frac{R_t^s e^{-R_t}}{s!} \cdot \frac{((N_t - 1)R_t)^{N_{t+1} - s}}{(N_{t+1} - s)!}}{\displaystyle\frac{(N_t R_t)^{N_{t+1}} e^{-N_t R_t}}{N_{t+1}!}} \nonumber\\
			& = \binom{N_{t+1}}{s} \left( \frac{1}{N_t} \right)^s \left( 1 - \frac{1}{N_t} \right)^{N_{t+1} - s}
	\end{align}
}

This is the probability mass function of a Binomial distribution and therefore we deduce that:
	\begin{equation}
		S_{t,1} \bigg| \left(\sum_{i=1}^{N_t} S_{t,i} = N_{t+1}\right) \sim \text{Binomial}\left(N_{t+1}, \frac{1}{N_t}\right)\label{eq:binomial}
	\end{equation}

The $k$-th falling factorial moments of $X \sim \mathrm{Binomial}(n,p)$ are
\citep{Potts1953}:

\begin{equation}
	\mathbb{E}\left[\frac{X!}{(X-k)!}\right]=\binom{n}{k} p^k k!
\end{equation}

By applying this formula to the Binomial distribution in Equation \eqref{eq:binomial}
and injecting into Equation \eqref{eq:GeneralInclusiveProb}, we deduce that the inclusive probability of coalescence for $k$ lines is:

\begin{equation}
	p_{k,t}(N_t, N_{t+1})=\frac{1}{N_t^{k-1}}\label{eq:PoisInclusiveProb}
\end{equation}

In addition, following a similar reasoning as for Equation \eqref{eq:binomial} we can show that:
\begin{equation}
X_{t,1} \bigg| \left(\sum_{i=1}^{N_t} X_{t,i} = n\right) \sim \text{Binomial}\left(n, \frac{1}{N_t}\right)
\end{equation}

By injecting the probability mass function of this Binomial distribution 
into Equation \eqref{eq:GeneralExclusiveProb} we deduce that 
the exclusive probability of coalescence for $k$ lines from a sample of $n$ $(n \geq k)$ is:
	\begin{equation}
		p_{n,k,t}(N_t, N_{t+1}) = \frac{(N_t-1)^{n-k}}{N_t^{n-1}}\label{eq:PoisExclusiveProb}
	\end{equation}
	
\begin{figure}[!p]
\begin{center}
\includegraphics[width=15cm]{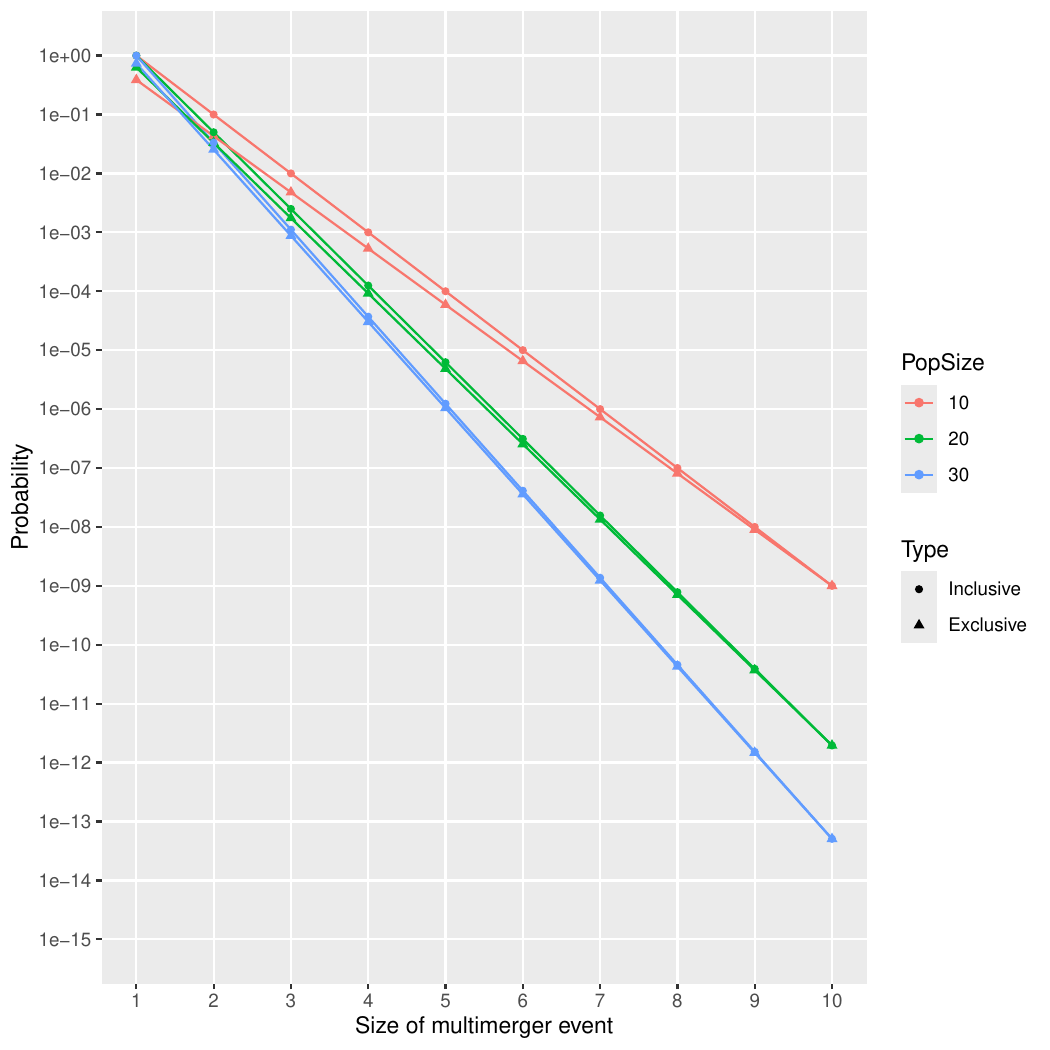}
\end{center}
\caption{Inclusive and exclusive coalescence probabilities for the Poisson case.
\label{fig:pois}}
\end{figure}

It is interesting to note that neither the inclusive nor the exclusive coalescence probability
depend on the mean $R_t$ of the Poisson offspring distribution or the size $N_{t+1}$
of the population at time $t+1$. Both only depend on the population size $N_t$ at time $t$.
The inclusive coalescent probability in Equation \eqref{eq:PoisInclusiveProb}
can also be obtained conceptually by considering
that among the $k$ lines, the first one has an ancestor with probability one, 
and the remaining $k-1$ need to have the same ancestor among a set of $N_t$ from
which they choose uniformly at random so that the probability of picking the same ancestor
is $1/N_t$. The exclusive coalescent probability in Equation \eqref{eq:PoisExclusiveProb} 
can be derived likewise by considering
that in addition to the above, each of the $n-k$ other lines need to choose a different
ancestor, which happens with probability $(N_t-1)/N_t$.
Figure \ref{fig:pois} illustrates the inclusive and exclusive coalescence probabilities 
for a set of size $k=1$ to $k=10$ amongst a total of $n=10$ observed
individuals, in a population of size $N_t=10$, $N_t=20$ or $N_t=30$. 

\section{Negative-Binomial offspring distribution case}

In this section we consider that the offspring distribution is 
Negative-Binomial, a distribution often used to model superspreading individuals
\citep{Lloyd-Smith2005} and which can also be used to model superspreading events 
\citep{Craddock2025}.
Let $\alpha_t=\text{Negative-Binomial}(r,p)$ with parameters $(r,p)$ set by moment-matching the mean $R_t$ and variance $V_t$ of the offspring distribution
which are assumed constant over time.
The resulting parameters for this distribution are $r=R_t^2/(V_t-R_t)$ and $p=R_t/V_t$.
In this case, we have:
	\begin{equation}
		\sum_{i=1}^{N_t} S_{t,i} \sim \text{Negative-Binomial}(N_t r,p)
	\end{equation}
and similarly to the Poisson offspring distribution case we identify that the conditional distribution of $S_{t,1} | \sum_{i=1}^{N_t} S_{t,i}$ is as follows:

{\allowdisplaybreaks
	\begin{align}
		\mathbb{P}\bigg[ S_{t,1} = s \bigg| \sum_{i=1}^{N_t} S_{t,i} = N_{t+1} \bigg]
			& = \frac{\alpha_t(s) \cdot \mathbb{P} \bigg[\sum_{i=2}^{N_t} S_{t,i} = N_{t+1} - s \bigg]}{\mathbb{P}\bigg[\sum_{i=1}^{N_t} S_{t,i} = N_{t+1} \bigg]} \nonumber\\
			& = \frac{\displaystyle\frac{\Gamma(r+s)}{s! \Gamma(r)} (1-p)^s p^r \cdot \frac{\Gamma\big((N_t - 1)r+(N_{t+1} - s)\big)}{(N_{t+1} - s)! \Gamma((N_t - 1) r)} (1-p)^{N_{t+1} - s} p^{(N_t-1)r}}{\displaystyle\frac{\Gamma(N_tr+N_{t+1})}{N_{t+1}! \Gamma(N_t r)} (1-p)^{N_{t+1}} p^{N_t r}} \nonumber\\
			& = \frac{N_{t+1}!}{s! (N_{t+1} - s)!} \frac{\Gamma(r+s) \Gamma\big( (N_t - 1)r + (N_{t+1} - s) \big)}{\Gamma(N_t r + N_{t+1})} \frac{\Gamma(N_t r)}{\Gamma(r) \Gamma\big((N_t - 1)r \big)} \nonumber\\
			& = \binom{N_{t+1}}{s} \frac{\mathrm{B}(s + r, N_{t+1} - s + (N_t-1)r)}{\mathrm{B}(r, (N_t-1)r)}\label{eq:derivBetaBinom}
	\end{align}
}

\noindent where  $\mathrm{B}(x,y)$ denotes the Beta function defined as $\mathrm{B}(x,y)=\Gamma(x)\Gamma(y)/\Gamma(x+y)$. This is the probability mass function of a Beta-Binomial distribution and therefore we deduce that:

\begin{equation}
S_{t,1} \bigg| \bigg(\sum_{i=1}^{N_t} S_{t,i} = N_{t+1} \bigg) \sim \text{Beta-Binomial}(N_{t+1},r, (N_t - 1)r)
\label{eq:beta-binomial}
\end{equation}

The $k$-th falling factorial moments of $X \sim \text{Beta-Binomial}(n,\alpha,\beta)$ are \citep{Tripathi1994}:

\begin{equation}
	\mathbb{E}\left[\frac{X!}{(X-k)!}\right]=\binom{n}{k} \frac{\mathrm{B}(\alpha+k,\beta)k!}{\mathrm{B}(\alpha,\beta)}
\end{equation}

By applying this formula to the Beta-Binomial distribution in Equation \eqref{eq:beta-binomial}
and injecting into Equation \eqref{eq:GeneralInclusiveProb}, we deduce that the inclusive probability of coalescence for $k$ lines is:

\begin{equation}
p_{k,t}(N_t, N_{t+1})=
\frac{\mathrm{B}(N_t r+1,r+k)}{\mathrm{B}(r+1,N_t r+k)}
\label{eq:NegBinInclusiveProb}
\end{equation}

In addition, following a similar reasoning as for
Equation \eqref{eq:beta-binomial}, we can show that:

\begin{equation}
X_{t,1} \bigg| \bigg(\sum_{i=1}^{N_t} X_{t,i} = n \bigg) \sim \text{Beta-Binomial}(n,r, (N_t - 1)r)
\end{equation}

By injecting the probability mass function of this Beta-Binomial distribution into Equation 
\eqref{eq:GeneralExclusiveProb} we deduce that 
the exclusive probability of coalescence for $k$ lines is:

\begin{equation}
p_{n,k,t}(N_t, N_{t+1})=\frac{N_t \mathrm{B}(k+r, n-k+N_t r-r)}{\mathrm{B}(r, N_t r-r)}
\label{eq:NegBinExclusiveProb}
\end{equation}

It is interesting to note that as for the Poisson case, the inclusive and exclusive coalescence probabilities do not depend on the size $N_{t+1}$ of the population at time $t+1$.
They both depend on the 
Negative-Binomial offspring distribution only through the dispersion parameter $r$.
If we consider that $r$ is large in Equations \eqref{eq:NegBinInclusiveProb} and
\eqref{eq:NegBinExclusiveProb}, we can 
derive that the asymptotic behaviour is the same as in the Poisson case  
shown in Equations  \eqref{eq:PoisInclusiveProb} and
\eqref{eq:PoisExclusiveProb}. For example this can be derived by rewriting the
Beta functions using Gamma functions, and using the following form of Stirling's approximation:
\begin{equation}
\mathrm{lim}_{a \rightarrow \infty} \frac{\Gamma(a+b)}{\Gamma(a)}=a^b \mathrm{e}^{-b}
\label{eq:stirling}
\end{equation}

\begin{figure}[!p]
\begin{center}
\includegraphics[width=15cm]{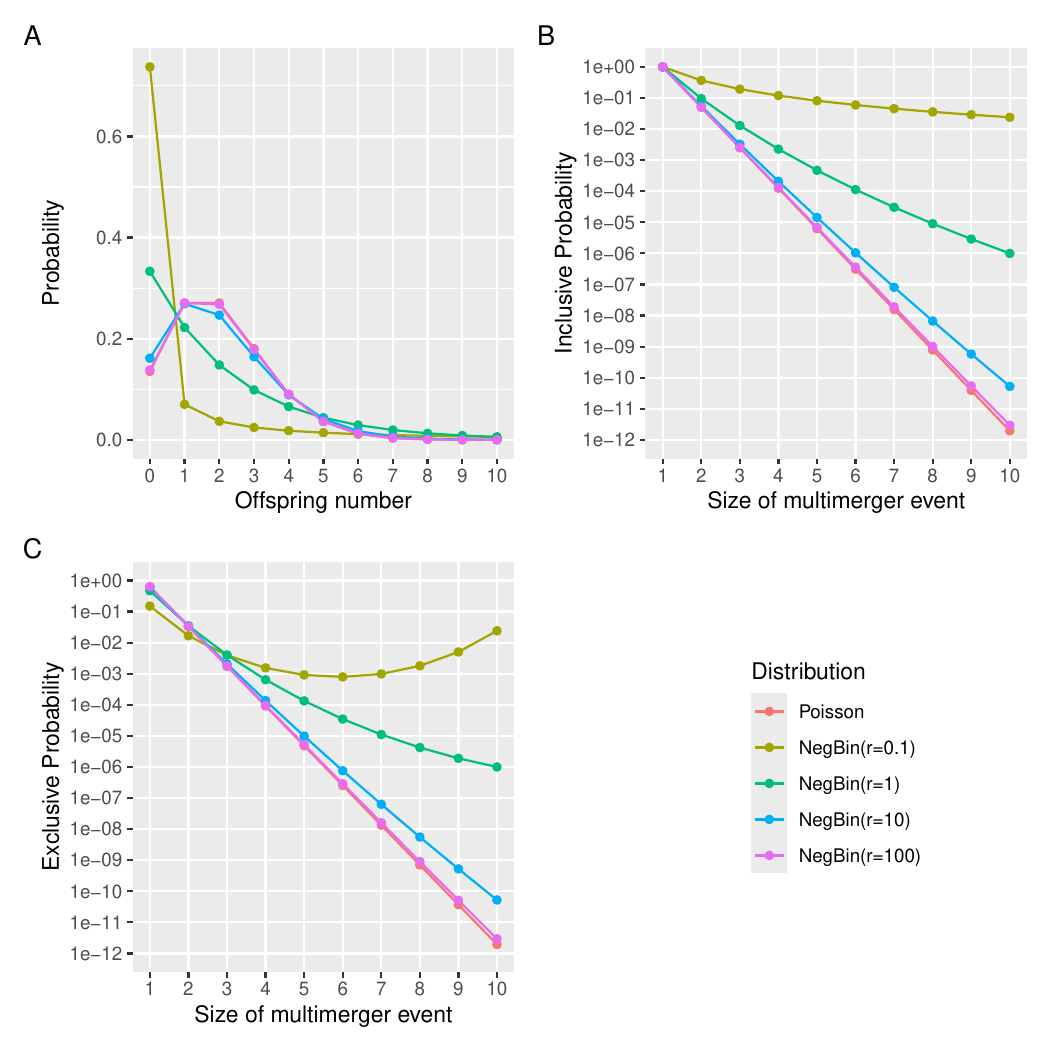}
\end{center}
\caption{(A) Offspring distributions with mean $R_t=2$. (B) Inclusive probability of coalescence for $N_t=20$ and $n=10$. (C) Exclusive probability of coalescence for $N_t=20$ and $n=10$.
\label{fig:negbin}}
\end{figure}

Figure \ref{fig:negbin} illustrates the inclusive and exclusive coalescence probabilities for
the Negative-Binomial case for a set of size $k=1$ to $k=10$ amongst a total of $n=10$
 observed lines, in a population with size $N_t=20$. 
Several Negative-Binomial offspring distributions are compared, 
all of which have the same mean $R_t=2$,
and with the dispersion parameter equal to $r=0.1$, $r=1$, $r=10$ and $r=100$
(Figure \ref{fig:negbin}A).
When $r=1$ the Negative-Binomial reduces to a Geometric distribution.
When $r$ is high 
the dispersion is low and the Negative-Binomial case 
behaves almost like the Poisson case
for both the inclusive (Figure \ref{fig:negbin}B) and the exclusive coalescence
probabilities (Figure \ref{fig:negbin}C).
When $r$ is lower the dispersion of the
offspring distribution increases, so that 
both the inclusive and exclusive probabilities of larger multiple merger events are increased
compared to the Poisson case.
In particular, when $r=0.1$ we see that the exclusive probability can increase
with the size of the event considered (Figure \ref{fig:negbin}C). This happens because
the probability is not much lower for the common ancestor having say 10 rather than 9 offspring,
while on the other hand if the event is of size 9 only then another individual in the generation
of the ancestor needs to have had at least one sampled offspring. 

\section{Limit when the population size is large}

Here we consider that the population size $N_t$ is fixed and large, so that 
we can show the connections
between our results and several previous studies on the ancestral process of infectious diseases.
In the Poisson case, from Equations \eqref{eq:PoisInclusiveProb} and \eqref{eq:PoisExclusiveProb} we can see that both inclusive and exclusive probabilities
are of order $\mathcal{O}(N_t^{1-k})$. We can therefore ignore events with $k>2$
and retain only the events with $k=2$, which means that there are only binary coalescent events
and no multiple merger events.
The binary coalescent events occur with the same inclusive and exclusive
probabilities:
\begin{equation}
p_{2,t}(N_t, N_{t+1})=p_{n,2,t}(N_t, N_{t+1})=\frac{1}{N_t}\label{eq:poissonapprox}
\end{equation}

For the Negative-Binomial case, from Equations \eqref{eq:NegBinInclusiveProb} and \eqref{eq:NegBinExclusiveProb} we can rewrite using Gamma functions and apply the 
form of Stirling's equation given in Equation \eqref{eq:stirling} to show that 
once again both inclusive and exclusive probabilities are also of order $\mathcal{O}(N_t^{1-k})$. 
We can therefore once again ignore events with $k>2$
and retain only the events with $k=2$ which occur with the same inclusive and 
exclusive probabilities:

\begin{equation}
p_{2,t}(N_t, N_{t+1})=p_{n,2,t}(N_t, N_{t+1})=\frac{r+1}{N_t r +1} \approx \frac{r+1}{N_t r}\label{eq:negbinapprox}
\end{equation}

\citet{koelleRatesCoalescenceCommon2012} derived the rates of coalescence of two lineages 
for several epidemiological models, assuming a large population at equilibrium.
For each model they use the equation $N_\mathrm{e}=N/\sigma^2$ to relate the effective population
size $N_\mathrm{e}$ to the actual population size $N$ and the variance $\sigma^2$ in the number
of offspring. This relationship was first established by \citet{Kingman1982} to
derive the backward-in-time coalescent model from the forward-in-time
Cannings exchangeable models \citep{Cannings1974}. 
This result implies that the rate of coalescence for two lineages is 
$1/N_\mathrm{e}=\sigma^2/N$.
From Equation \eqref{eq:negbinapprox} we can take $R_t=1$ to achieve 
equilibrium of the population size and the method of moments estimator
$r=R_t^2/(V_t-R_t)=1/(V_t-1)$ to deduce 
the equivalent result $p_{2,t}(N_t,N_{t+1})=V_t/N_t$. 

\citet{Volz2012a} showed that the rate of coalescence for two lineages under a continuous-time epidemic coalescent model is $2f(t)/I(t)^2$
 where $f(t)$ is the incidence of the disease and $I(t)$ its prevalence. 
 Setting in this formula the prevalence as $I(t)=N_{t+1}=N_t R_t$ and the incidence as
 $f(t)=R_t I(t)=R_t^2 N_t$ we get a coalescent rate of 
 $2/N_t$. To apply our methodology we need to consider that
the offspring distribution is Geometric, since the epidemiological
 models considered have successes (transmission) happening until the first failure
 (removal). We therefore set $r=1$ in Equation \eqref{eq:negbinapprox} 
 to make the Negative-Binomial offspring distribution
 reduce to a Geometric distribution and the same result follows.

\citet{Fraser2017} calculated the effective population size $N_\mathrm{e}(t)$ 
as a function of the actual population size $N(t)$ and the mean and variance of the offspring distribution $R$ and $\sigma^2$. 
This formula was used to estimate the dispersion parameter of a Negative-Binomial offspring distribution from genetic data \citep{Li2017}.
Using our notations, their formula is equivalent to the inclusive coalescence 
probability for two lineages:

\begin{equation}
 p_{2,t}(N_t, N_{t+1})=\frac{V_t/R_t+R_t-1}{N_t R_t}\label{eq:fraser}
 \end{equation}

In the Poisson case we have $V_t=R_t$ so that Equation \eqref{eq:fraser} simplifies to $1/N_t$ which agrees with Equation \eqref{eq:poissonapprox}. 
In the Negative-Binomial case we have $V_t/R_t=1/p=1+R_t/r$ so that Equation \eqref{eq:fraser}
simplifies to $(r+1)/(N_t r)$ which agrees with our Equation \eqref{eq:negbinapprox}. 
Conversely, if we substitute the method of moments estimator 
$r=R_t^2/(V_t-R_t)$
in Equation \eqref{eq:negbinapprox} we obtain the Equation \eqref{eq:fraser}
originally from \citet{Fraser2017}. 

\section{Definition of a new lambda-coalescent model}

The coalescent model \citep{Kingman1982,Kingman1982a} describes the ancestry of a sample from a large population 
evolving according to many forward-in-time models such as the Wright-Fisher model \citep{Wright1931,Fisher1930}, 
the Moran model \citep{Moran1958} and the Cannings exchangeable model \citep{Cannings1974}.
Since the coalescent considers a large population in which each individual only has a number of offspring that is small
compared to the population size, coalescent trees are always binary and do not feature multiple mergers,
making them unsuitable to represent the ancestry of outbreaks considered in this study.
However, the lambda-coalescent model is an extension of the coalescent model that 
allows multiple mergers
\citep{pitmanCoalescentsMultipleCollisions1999,sagitovGeneralCoalescentAsynchronous1999,donnellyParticleRepresentationsMeasureValued1999}. 

A lambda-coalescent model is defined by a probability measure 
$\Lambda(\mathrm{d} x)$ on the interval $[0,1]$, from which we deduce
the rate $\lambda_{n,k}$ at which any subset of $k$ lineages within a set of $n$ observed lineages 
coalesce:

\begin{equation}
    \lambda_{n,k} = \int_{0}^{1}{x^{k-2}(1-x)^{n-k}\,\Lambda(\mathrm{d} x)}\label{eq:lambda}
\end{equation}

The beta-coalescent \citep{schweinsbergCoalescentProcessesObtained2003} is a 
specific type of lambda-coalescent that has been used recently in several studies
analysing genetic data from infectious disease agents \citep{Hoscheit2019,Menardo2021,Helekal2024,zhangMultipleMergerCoalescent2024}.
The beta-coalescent model 
has a single parameter $\alpha \in [0,2]$ and is defined as:

\begin{equation}
\Lambda(\mathrm{d}x)=\frac{x^{1-\alpha}(1-x)^{\alpha-1}}{\mathrm{B}(2-\alpha,\alpha)}\mathrm{d}x
\label{eq:beta0}
\end{equation}

By combining Equations \eqref{eq:lambda} and \eqref{eq:beta0} we deduce that:

\begin{equation}
\lambda_{n,k}=\frac{\mathrm{B}(k-\alpha,n-k+\alpha)}{\mathrm{B}(2-\alpha,\alpha)}
\label{eq:beta}
\end{equation}

Special cases of the beta-coalescent 
include $\alpha=2$ corresponding to the Kingman coalescent,
$\alpha=1$ which is known as the Bolthausen-Sznitman coalescent
and $\alpha=0$ for which the phylogeny is always star-shaped.

We now define a new lambda-coalescent based on the Negative-Binomial case described previously. We call this new lambda-coalescent model the omega-coalescent (where omega stands
for outbreak).
For ease of comparison with other coalescent models, we consider that time is continuous
and that the population size remains constant equal to $N_t=N$. 
The exclusive coalescent probability $p_{n,k,t}(N_t, N_{t+1})$ in the Negative-Binomial case 
given by Equation \eqref{eq:NegBinExclusiveProb} can be used to determine the corresponding 
rate of the  omega-coalescent, if we consider that the probability of each event in discrete time
is equal to the constant rate of this event happening in continuous time: 
\begin{equation}
\lambda_{n,k}=p_{n,k,t}(N_t=N, N_{t+1}=N)=\frac{N \mathrm{B}(k+r, n-k+N r-r)}{\mathrm{B}(r, N r-r)}
\label{eq:omega}
\end{equation}

Note that this equation implies that continuous time is measured
approximately in number of transmission generations. 
For example to measure time in decimal days instead, the time scale
would need to be multiplied by the mean of the generation time 
distribution measured in days \citep{Svensson2007}.

For a lambda-coalescent model to be consistent, 
when a multiple merger of size $k$ amongst $n$ lineages occurs, if 
an additional lineage is revealed it must either take part in the multiple merger or 
remain unaffected \citep{berestyckiRecentProgressCoalescent2009}.
This implies that the rates must satisfy: 
\begin{equation}
\lambda_{n,k}=\lambda_{n+1,k}+\lambda_{n+1,k+1}
\label{eq:consistency}
\end{equation}
This consistency property is easily verified for the beta-coalescent in Equation \eqref{eq:beta}
and likewise for the omega-coalescent in Equation \eqref{eq:omega}, in both
cases using recursive properties of the Beta functions used in the respective definitions.

The omega-coalescent has two parameters: the constant population size $N$ and
the dispersion parameter $r$. 
In order to compare the omega-coalescent defined in Equation \eqref{eq:omega}
with other models such as the beta-coalescent defined in Equation \eqref{eq:beta}, we consider 
the distribution of the size $k$ of the next event among
a set of $n$ lineages. For any lambda-coalescent this can be computed as:

\begin{equation}
p(k|n)=\frac{\binom{n}{k}\lambda_{n,k}}{\sum_{i=2}^n \binom{n}{i}\lambda_{n,i}}
\label{eq:nextsize}
\end{equation}

\begin{figure}[!p]
\begin{center}
\includegraphics[width=15cm]{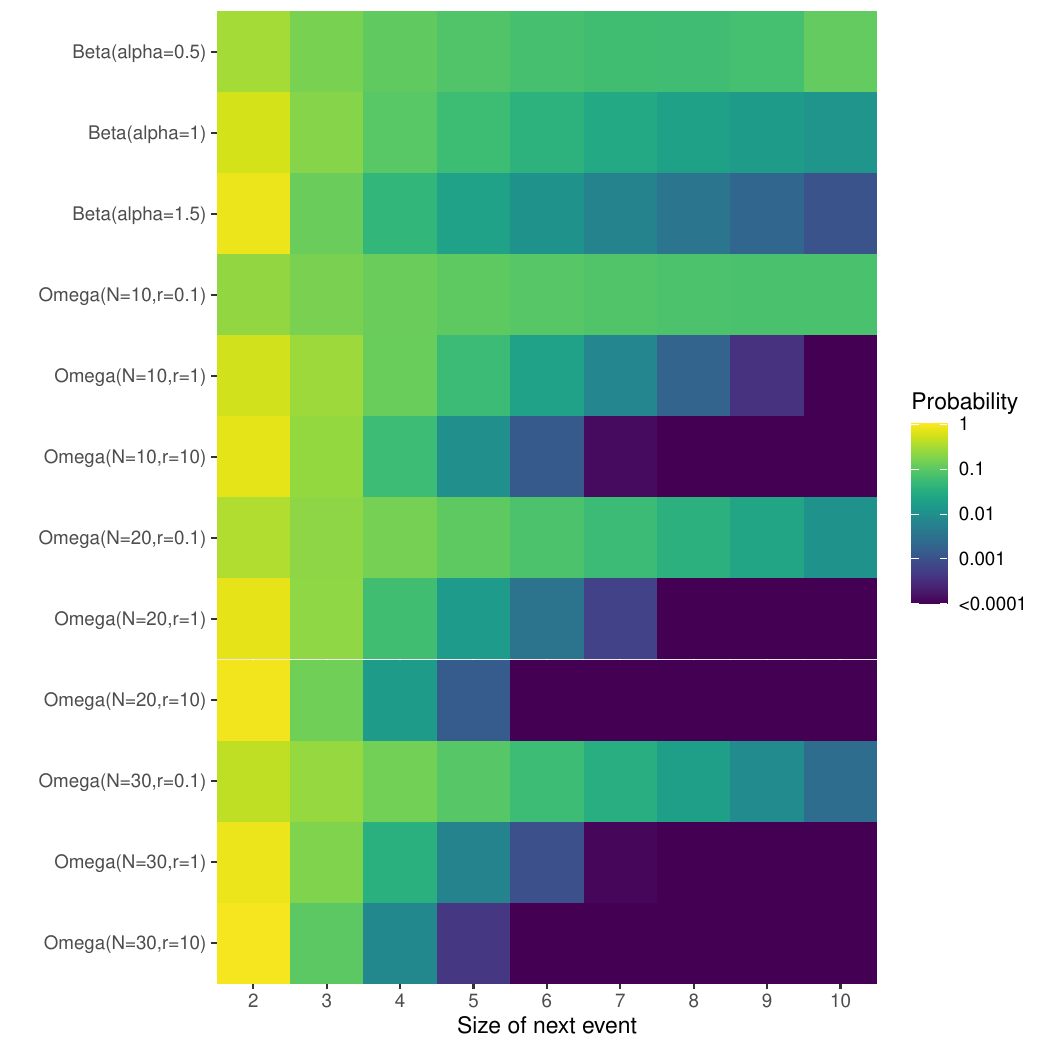}
\end{center}
\caption{Distribution of the size of the next event among a set of $n=10$ lineages, 
compared between the beta-coalescent and the omega-coalescent model 
with various parameters.
\label{fig:compare}}
\end{figure}

Figure \ref{fig:compare} compares this distribution for $n=10$ in the beta-coalescent with 
parameter $\alpha \in \{0.5,1,1.5\}$ and for the omega-coalescent with parameters
$N \in \{10,20,30\}$ and $r \in \{0.1,1,10\}$. In the beta-coalescent, the distribution
shifts towards more larger multiple merger events as the parameter $\alpha$ decreases.
In the omega-coalescent a wider range of behaviours is obtained when varying the two parameters
$N$ and $r$. For a given value of $N$, decreasing the value of $r$ results in more
larger events. Conversely, for a given value of $r$ we can see that increasing the
value of $N$ reduces the probability of larger events.

\begin{figure}[!p]
\begin{center}
\includegraphics[width=15cm]{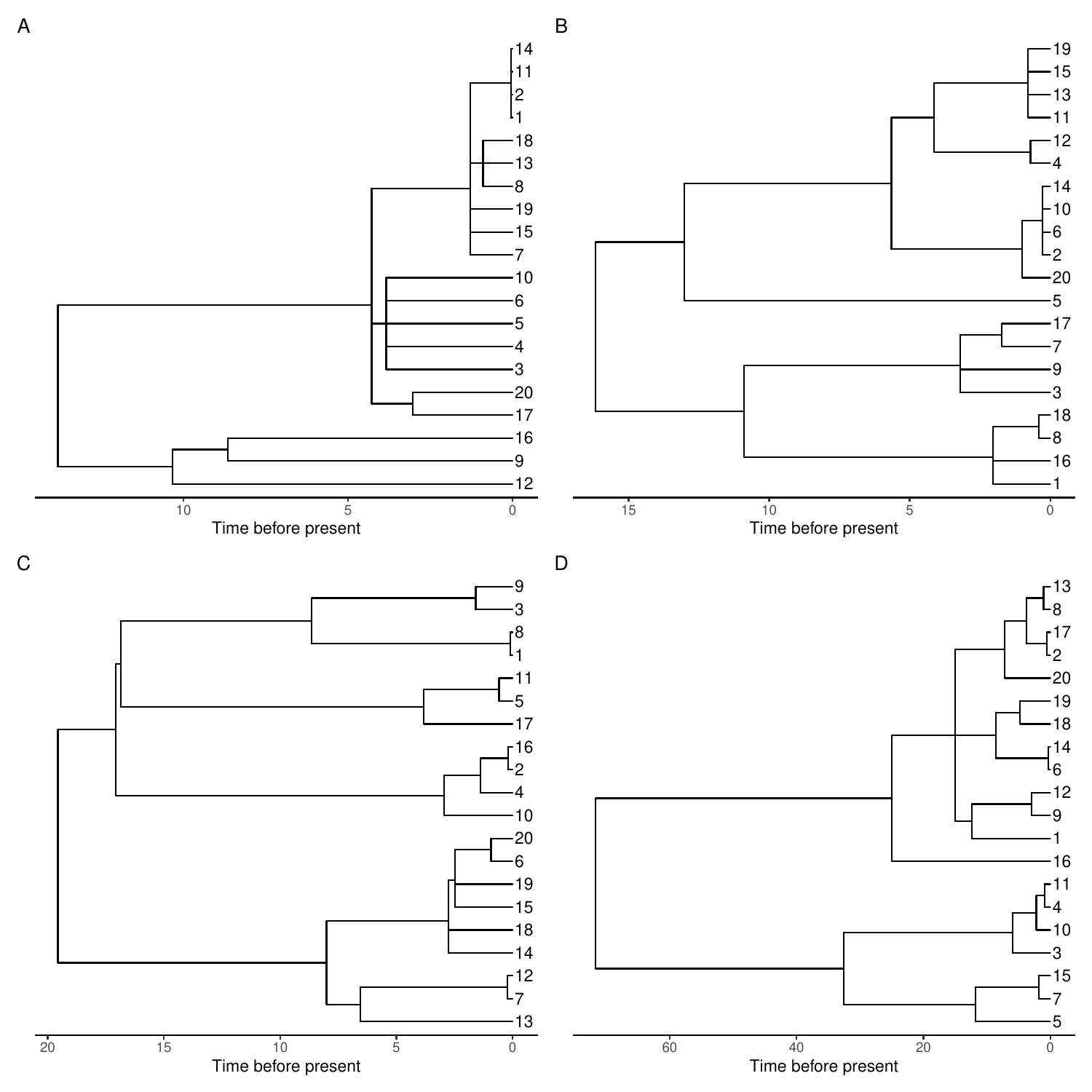}
\end{center}
\caption{Example of trees simulated under the omega-coalescent with $r=0.1$ (A),
$r=1$ (B), $r=10$ (C) and $r=100$ (D).
\label{fig:trees}}
\end{figure}

\begin{figure}[!p]
\begin{center}
\includegraphics[width=15cm]{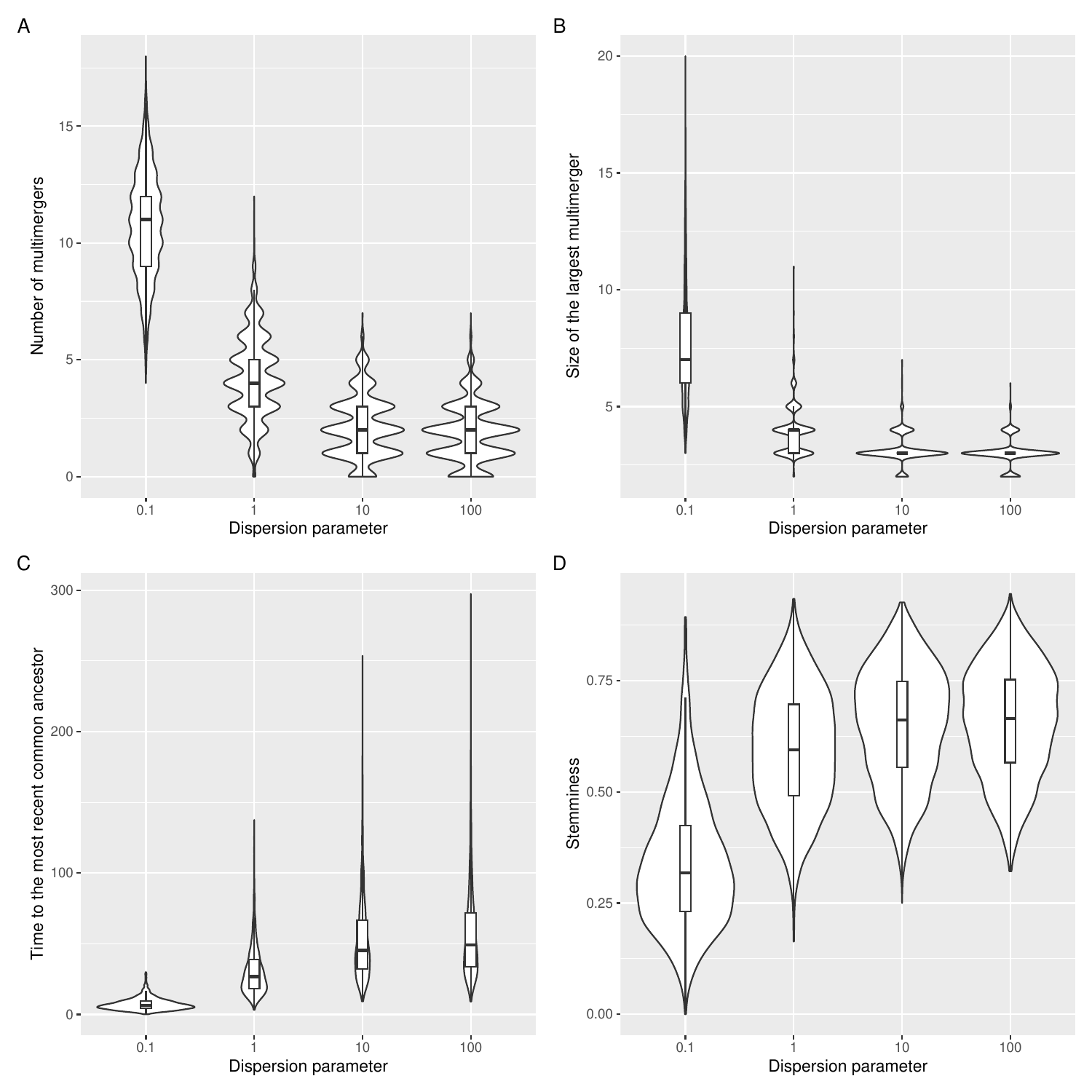}
\end{center}
\caption{Summary statistics for trees simulated under the omega-coalescent with $r=0.1$, $r=1$, $r=10$ and $r=100$, namely number of multiple mergers (A) the size of the largest multiple merger (B), the time to the most recent common ancestor (C) and the stemminess (D). 
\label{fig:stats}}
\end{figure}

Genealogies can be simulation from the omega-coalescent model 
defined in Equation \eqref{eq:omega} using the same algorithm
as for other lambda-coalescent models \citep{pitmanCoalescentsMultipleCollisions1999}.
Given $n$ lineages, the next coalescent event happens after a time that is 
exponentially distributed with rate $\sum_{i=2}^n \binom{n}{i}\lambda_{n,i}$,
the size $k$ of this event is drawn according to Equation \eqref{eq:nextsize},
and the $k$ lineages that coalesce are chosen uniformly amongst the $n$ lineages.
This process is repeated iteratively until all lineages have coalesced.
Figure \ref{fig:trees} shows examples of trees simulated for a sample of size $n=20$,
constant population size $N=30$ and dispersion parameter $r \in \{0.1,1,10,100\}$. 
It is already clear from these single realisations that the lower values of $r$ result
in trees with more larger multiple merger events and lower time to the most recent common ancestor,
but to quantify these properties we need to consider many trees.
Figure \ref{fig:stats} shows summary statistics for 10,000 trees simulated in the same conditions
as the individual trees shown in Figure \ref{fig:trees}. 
As the dispersion parameter increases from $r=0.1$
to $r=100$ multiple merger events become less and less likely and less large
(Figure \ref{fig:stats}A and B), and the 
time to the most recent common ancestor increases
(Figure \ref{fig:stats}C).
Furthermore, the stemminess of the tree increases, 
which is defined as the sum of lengths of internal branches divided by the total sum of branch lengths (Figure \ref{fig:stats}D). 
Stemminess is usually taken as a sign of population size dynamics
\citep{fialaFactorsDeterminingAccuracy1985,Didelot2009d},
which would be misleading here since all simulations assumed
a constant population size.

\section{Parameter inference}

Let us now consider a genealogy $T$ with $n$ leaves and $c$ coalescent nodes, with $t_0=0$ the
sampling time,
$t_1,...,t_c$ the times of the coalescent nodes in increasing order and $k_i$ the number
of lineages coalescing at time $t_i$. The number of lineages existing between
time $t_{i-1}$ and $t_i$ is then $n_i=n-\sum_{j=1}^{i-1} k_j$.
Under a lambda-coalescent model, the genealogy $T$ has likelihood:

\begin{equation}
p(T|\Lambda)=\prod_{i=1}^{c}\lambda_{n_i,k_i}\exp\left(-\sum_{j=2}^{n_i}\binom{n_i}{j}\lambda_{n_i,j}(t_i-t_{i-1})\right)
\label{eq:likelihood}
\end{equation}

Note that in Equation \eqref{eq:likelihood} the term $\binom{n_i}{k_i}$  term from the
coalescent rate cancels out with its reciprocal from the probability of sampling $k_i$ specific
lineages to coalesce within a set of $n_i$.
Estimating the lambda measure from Equation \eqref{eq:lambda} 
in general is a difficult problem
\citep{Koskela2018a,miropinaEstimatingLambdaMeasure2023}.
Here however we focus on estimation under the omega-coalescent model,
where the $\lambda_{n,k}$ terms are given by Equation \eqref{eq:omega}.
There are therefore two parameters to estimate which have direct and important 
biological meaning:
the effective population size $N$ (which remains constant) and the dispersion
parameter $r$ of the Negative-Binomial offspring distribution. 
We perform estimation simply by maximising the likelihood in Equation \eqref{eq:likelihood},
using the Brent algorithm \citep{brent1971algorithm} when estimating a single parameter
and the L-BFGS-B algorithm \citep{byrd1995limited} when estimating both parameters.

\begin{figure}[!p]
\begin{center}
\includegraphics[width=13cm]{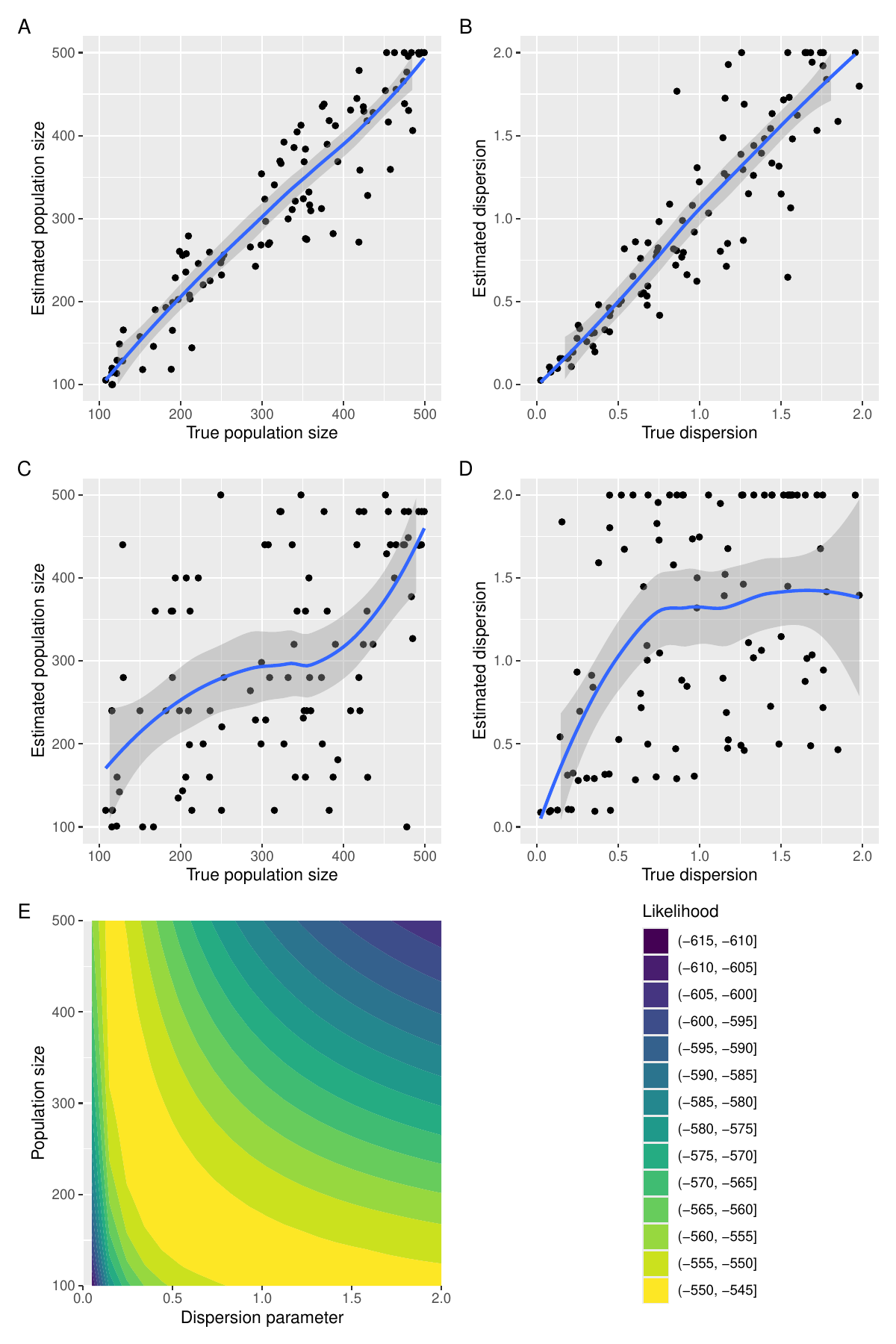}
\end{center}
\caption{Maximum likelihood estimation of parameters. (A) Estimation of the population
size given the dispersion parameter. (B) Estimation of the dispersion parameter
given the population size. (C and D) Joint estimation of both the population
size and dispersion parameters. (E) Example of likelihood surface as a function
of both parameters.
\label{fig:estim}}
\end{figure}

We simulated 100 genealogies from the omega-coalescent model each of which had
$n=100$ leaves, with parameter $N$ drawn uniformly at random between 100 and 500
and parameter $r$ drawn uniformly at random between 0.01 and 2.
If we assume knowledge of the dispersion parameter, then estimating the population size 
works really well (Figure \ref{fig:estim}A). Conversely we obtain good result when 
estimating the dispersion parameter given a known population size 
(Figure \ref{fig:estim}B). However, attempting to estimate both parameters at the same
time performed significantly less well (Figures \ref{fig:estim}C and D). 
To illustrate the cause of this, we consider a simulation for which the true 
parameters were $N=200$ and $r=0.5$, 
and we construct the likelihood surface (Figure \ref{fig:estim}E).
This shows a strong inverse tradeoff between the two parameters, which is
why it is harder to infer both parameters jointly.
This poor identifiability is analogous to the situation of a large population following 
the Cannings models \citep{Cannings1974}. In this case the 
coalescent process is fully determined by the effective population size 
$N_\mathrm{e}=N/\sigma^2$ as previously noted \citep{Kingman1982}, where $N$ is the 
population size and $\sigma^2$ is the variance in the number of offspring. Consequently
there is a full tradeoff between $N$ and $\sigma^2$, so that the ratio $N_\mathrm{e}$
can be estimated but not the parameters $N$ and $\sigma^2$ separately.

\section{Implementation}

We implemented the analytical methods described in this paper in a 
new R package entitled \emph{EpiLambda} which is available
at \url{https://github.com/xavierdidelot/EpiLambda} for R version 3.5 or later. 
All code and data needed to replicate the results are included in the ``run'' directory of the \emph{EpiLambda} repository.
The R package \verb+ape+ was used to store, manipulate and visualise phylogenetic trees
\citep{Paradis2019}.

\section{Discussion}

We have described an ancestral process for infectious diseases
which is relevant to the analysis of outbreaks of a relatively small size,
and to diseases with transmission heterogeneity.
We have shown how this process can be incorporated into a new lambda-coalescent
which we called the omega-coalescent. We only considered the
situation where all samples are taken at the same time, but
the omega-coalescent could be extended to allow temporally offset leaves 
following similar work on the coalescent \citep{Drummond2003} and the beta-coalescent \citep{Hoscheit2019}.
We also made the simplifying assumption of a constant population size,
but this could be relaxed 
following the same approach as previously described for integrating variable population size into
the coalescent \citep{Griffiths1994,Pybus2000,Ho2011skylinereview} and the beta-coalescent \citep{Hoscheit2019,zhangMultipleMergerCoalescent2024}.
Allowing the population size to vary could be especially useful for the omega-coalescent
for several reasons. Firstly, since it is aimed at relatively small outbreaks, it is likely that
their sizes varies significantly. Secondly, 
the probability of multiple merger events of various sizes
depends explicitly on the population size in Equation \eqref{eq:NegBinExclusiveProb}.
Changes in population size will therefore have an effect on the distribution of
events observed, as can be seen for example in Figure \ref{fig:compare}.
Thirdly, joint inference of a varying population size could help break the otherwise
difficult joint inference of a fixed population size with the dispersion parameter 
(Figure \ref{fig:estim}).

We compared the omega-coalescent only to the beta-coalescent \citep{schweinsbergCoalescentProcessesObtained2003} in Figure \ref{fig:compare}
as it is the model that has been most frequently used for infectious diseases
\citep{Hoscheit2019,Menardo2021,Helekal2024}.
Several other lambda-coalescent models have been proposed previously, such as 
the Dirac coalescent \citep{eldonCoalescentProcessesWhen2006},
the Durrett-Schweinsberg coalescent \citep{Durrett2005a} or the
extended Beta-coalescent \citep{Helekal2024}. However, none of these models
is equivalent to the omega-coalescent model. Indeed these previously
described lambda-coalescent models are mostly
concerned with situations where an individual can be the father of a significant
portion of a population in spite of the population being large, as opposed to the 
small populations with superspreading we considered here.
The xi-coalescent models are extensions to the lambda-coalescent models that admit multiple 
simultaneous mergers \citep{schweinsbergCoalescentsSimultaneousMultiple2000}.
This is clearly relevant to our basic discrete time model for small outbreaks,
since in small populations it is quite likely that separate subsets of individuals
have the same infector in the previous generation. 
However the exact timing of ancestry events is never available so that we
must rely on ancestral dating estimation with no notion of event
co-occurrence \citep{Volz2017,Didelot2018,Bouckaert2019,Helekal2024}. 
We therefore introduced a 
continuous time approximation in Equation \eqref{eq:omega} so that ancestry events 
do not co-occur.
The exact coalescent process for the discrete time Wright-Fisher process in a small 
population has been previously described \citep{fu_exact_2006}. 
It would be difficult however to extend this approach to the more complex
forward-in-time model we considered here, with variable population size and specific offspring
distribution, which further justifies our continuous time approximation. 

Finally, it should be noted that our model describes the transmission tree during
an outbreak, which is different from a phylogeny \citep{Jombart2011}. 
This difference is often ignored and in some settings it might be appropriate to do so,
but not always. Consequently, some
previous studies have used models of within-host evolution to bridge
the gap between transmission and phylogenetic trees
\citep{Didelot2014,Hall2015,Didelot2017}. However, these models assume
that each transmission event happens independently from one infector
to each of its infectees. This is not necessarily true especially when considering
superspreading events in which many individuals can become infected
simultaneously \citep{Riley2003,Wallinga2004,hoAccountingPotentialOverdispersion2023,Craddock2025}.
In conclusion, we have described a new ancestral model for infectious disease
outbreaks, which we hope will be useful especially in settings where 
the outbreaks are small or in the presence of high transmission heterogeneity.

\section*{Acknowledgements}

We acknowledge funding from the National Institute for Health Research (NIHR) Health Protection Research Unit in Genomics and Enabling Data.

\clearpage
\bibliographystyle{elsarticle-harv}
\bibliography{biblio}

\end{document}